\documentclass[prl,twocolumn,a4paper,amsmath,floatfix]{revtex4}
\usepackage[utf8]{inputenc}
\usepackage{graphicx}
\usepackage{amsmath}

\begin{document}
\title{Comment on ``The elusive fluid-and-crystal coexistence state in simulations
of monodisperse, hard-sphere colloids''}

\author{Frank Smallenburg$^1$}
\affiliation{$^1$Universit\'e Paris-Saclay, CNRS, Laboratoire de Physique des Solides, 91405 Orsay, France}

\begin{abstract}
In a recent article \cite{wang2026elusive}, Wang \textit{et al.} discuss the absence of simulations of monodisperse hard spheres in which a metastable fluid spontaneously nucleates into a stable fluid-crystal coexistence. Here, we show that such a simulation can be readily accomplished with standard simulation methods.
\end{abstract}
\maketitle

A recent article by Wang \textit{et al.}\cite{wang2026elusive} examines the ``elusive'' fluid-crystal coexistence in monodisperse hard spheres. Specifically, it points out that while many simulation studies have examined hard-sphere coexistence by deliberately initializing the system in a coexisting state, there are no reports of the observation of a \textit{spontaneous} crystal nucleation event resulting in a coexistence -- in practice, the crystal that grows out from spontaneous nucleation typically leads to a fully crystalline system.  The same article suggests that ``an unbiased simulation at coexistence would, on average, have to run for $3.17\times10^8$ years -- about 2\% of the age of the universe --
to observe a single spontaneous nucleation event that produces a stable fluid–crystal coexistence state.'' In this comment, we demonstrate that the actual simulations required are many orders of magnitude shorter.

There are two main issues in observing the spontaneous formation of a stable phase coexistence out of a homogeneous metastable fluid phase of hard spheres in simulations. First, the packing fraction ($\eta$) of the system needs to be high enough for spontaneous nucleation to occur on a feasible time scale. Based on recent simulation studies \cite{wohler2022hard, de2025local}, this puts the minimum packing fraction somewhere around $\eta_\mathrm{min} \gtrsim 0.53$. Second, the packing fraction needs to be low enough to avoid the entire system from crystallizing. In the thermodynamic limit of infinite system sizes, this puts the upper limit for the packing fraction at the melting point ($\eta_m \simeq 0.543$ \cite{smallenburg2024simple}). Within this region, the lever rule predicts what fraction $V_X/V$ of the total box volume $V$ is expected to be crystalline at global packing fraction $\eta$:
\begin{equation}
    \frac{V_X}{V} = \frac{\eta - \eta_f}{\eta_m -\eta_f}, \label{eq:lever}
\end{equation}
where the freezing packing fraction $\eta_f \simeq 0.4918$\cite{smallenburg2024simple}. Note that this based on $\eta_\mathrm{min}$, this implies that at least about three quarters of the simulation box is expected to be crystalline after spontaneous nucleation in a simulation. Importantly, for finite systems, the interface between the two phases will significantly affect the free energy of the system, and hence its final structure. For cubic simulation boxes, depending on the system size and relative volume of the two phases, the most stable state can be a slab-like geometry,  a system-spanning ``tube'' of the smaller phase surrounded by the larger one, a ''bubble'' of the smaller phase, or a homogeneous state (even when the system is slightly inside the coexistence regime) \cite{macdowell2006nucleation}. Larger system sizes will help to avoid the homogeneous state, by reducing the relative influence of the interfaces on the system. Moreover, larger system sizes will also make it easier to clearly observe distinct regions of fluid and crystal in the system.

\begin{figure}
    \raggedright
    a) \hfill\\
        \includegraphics[width=0.99\linewidth]{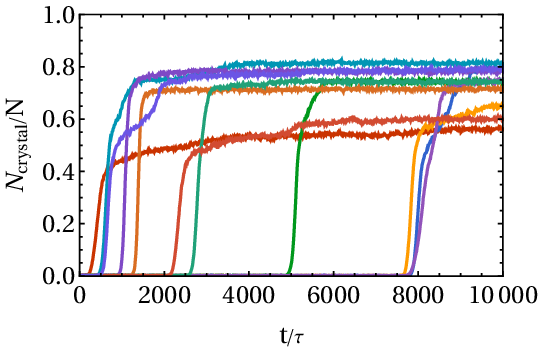}
    \begin{tabular}{ll}
    b)&c)\\
    \includegraphics[width=0.45\linewidth]{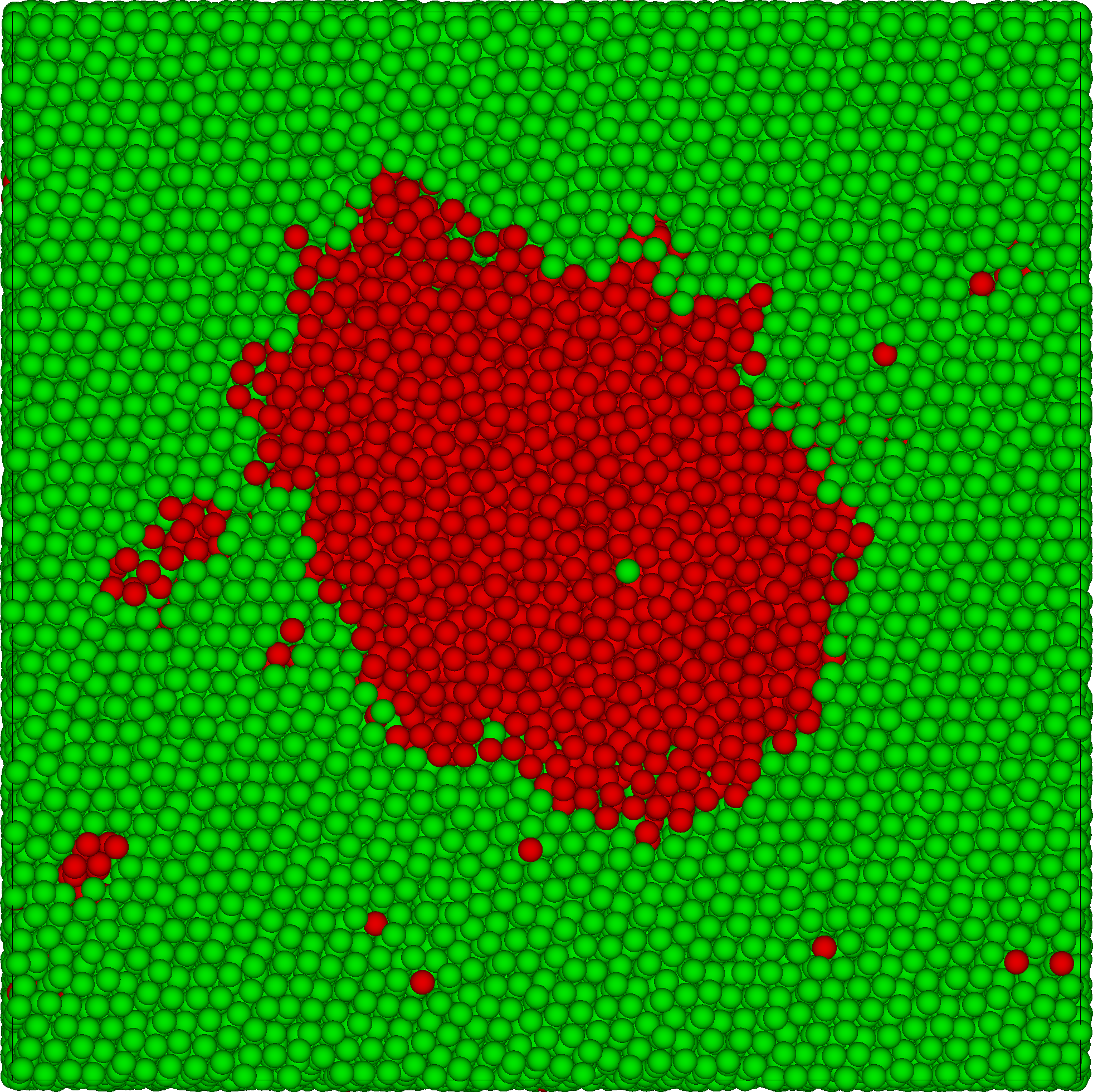}&
    \includegraphics[width=0.45\linewidth]{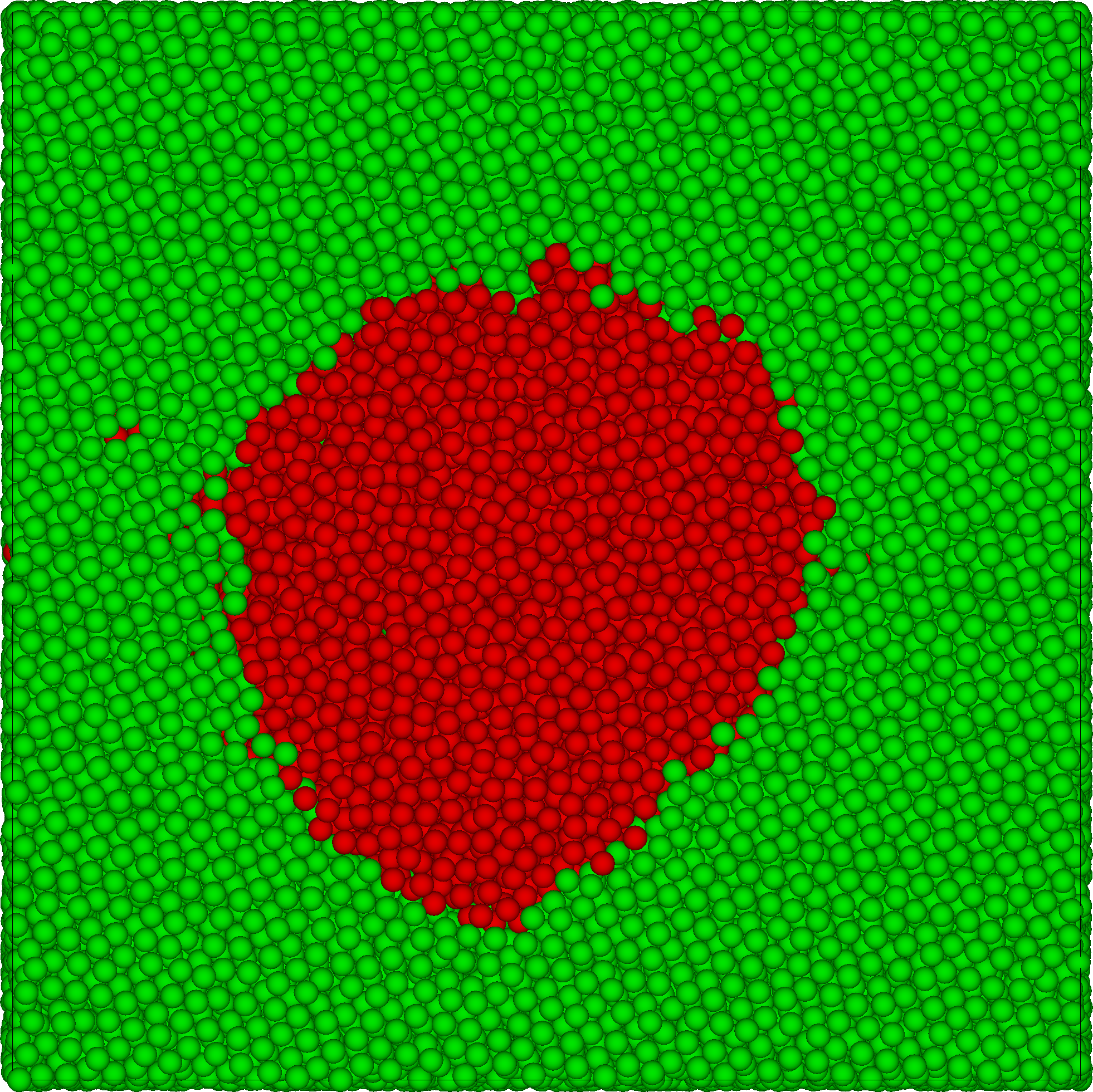}
    \end{tabular}
    \caption{a) Fraction of particles considered crystalline based on the criterion of having at least $7$ Ten Wolde bonds, as a function of time.  b,c) Snapshots of spontaneously formed fluid-crystal coexistence in a cubic box. Particles are colored as fluid-like (red) or crystal-like (green), based on the Lechner-Dellago averaged bond order parameter $\bar{q}_6$. Here, the fluid particles form a bubble within the surrounding crystal phase.     
    }
    \label{fig:bigbox}
\end{figure}

Hence, to observe spontaneously formed coexistence, we ideally want to perform simulations at a low packing fraction close to $\eta_m$ and in large systems. Of course, these two desires run counter to minimizing computational cost: lower packing fractions drastically increase the simulation time required for spontaneous nucleation, and larger system sizes slow down the overall simulation speed (although this is partially offset by having a larger volume to nucleate in). Nonetheless, using normal, unbiased event-driven molecular dynamics simulations \cite{smallenburg2022efficient}, observing spontaneous phase separation turns out to be quite doable for hard spheres.

We set the global packing fraction to $\eta = 0.5325$, and simulate 50 independent systems of $N=8 \times 10^4$ particles in a cubic simulation box. Each system is initialized in a disordered state, obtained by rapidly growing particles from a low-density configuration until the desired packing fraction is reached. We monitor the evolution of the largest nucleus with standard Ten Wolde bond order parameters \cite{ten1995numerical} \footnote{We use the solid-angle nearest-neighbor algorithm \cite{van2012parameter} to determine neighbors, use a minimal dot product value 0.7 for a crystalline bond, and consider particles to be crystalline if they have at least 7 crystalline bonds.}  We simulate each system for $10^4 \tau$ of simulation time, where $\tau = \sqrt{m \sigma^2/k_B T}$, $m$ is the particle mass, $k_B T$ is the thermal energy, and $\sigma$ is the particle diameter. During this time, 11 of the runs indeed spontaneously nucleate into a crystal, growing out to fill most of the system. 
In Fig. \ref{fig:bigbox}, we plot the evolution of the fraction of particles that are considered crystalline as a function of  as a function of simulation time. For each of these trajectories, there is a clear waiting time, consistent with spontaneous nucleation. The final degree of crystallinity varies per run, depending on the morphology of the interfaces and the presence of grain boundaries and other defects in the system. Two typical snapshots are shown in Figs. \ref{fig:bigbox}b and \ref{fig:bigbox}c, where we have colored the particles based on the value of the Lechner-Dellago averaged local bond order parameter \cite{lechner2008accurate} $\bar{q}_6$, which is less sensitive to crystal defects than the Ten Wolde bonds with our chosen cutoffs. In these systems, the remaining fluid particles typically form a bubble or tube within the surrounding crystal phase, as expected when most of the system is crystalline.

\begin{figure}
\raggedright
    \vspace{0.3cm}
    a)\\
    \includegraphics[width=0.99\linewidth]{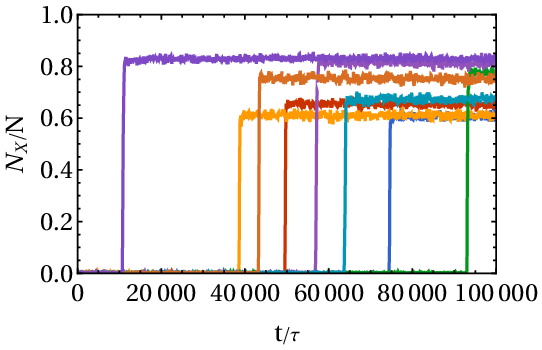}
    b)\\
    \includegraphics[width=0.99\linewidth]{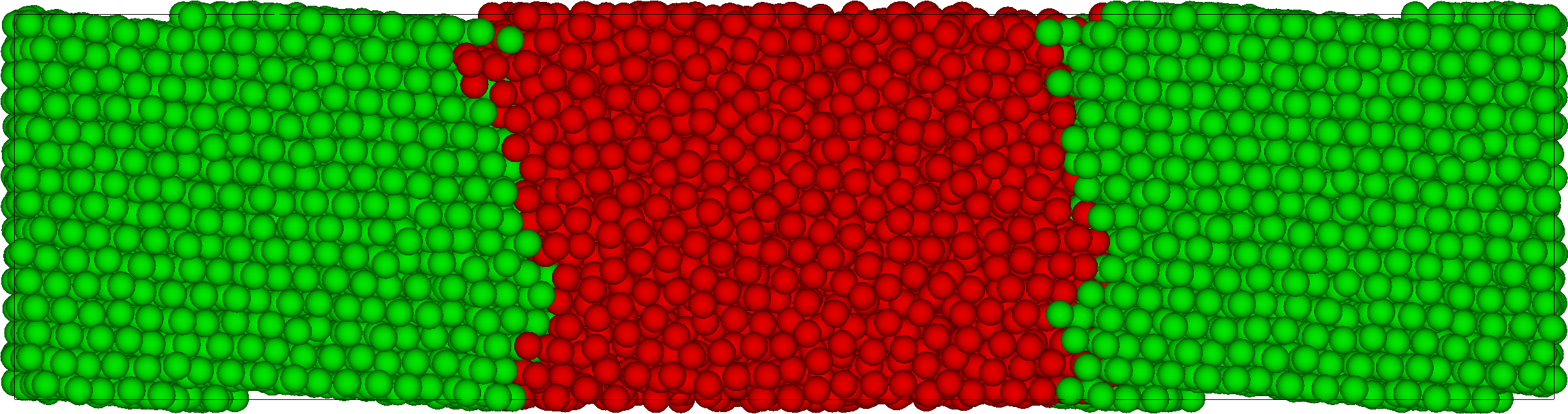}\\
    \caption{Same as Fig. \ref{fig:bigbox}, but for an elongated box containing $N=10^4$ particles. Note that simulation times are longer here, since smaller systems are faster to simulate.
    }
    \label{fig:longbox}
\end{figure}

If a slab-like geometry is desired, this can most easily be accomplished by performing the simulations in an elongated simulation box. To demonstrate this, we repeat the same simulations for a smaller system of $N=10^4$ particles, simulated for $10^5 \tau$ in a cuboidal box where one axis is 4 times as long as the other two. The corresponding cluster size evolutions for all 8 nucleating runs for this system (out of 50) are shown in Fig. \ref{fig:longbox}, along with a typical snapshot of the resulting coexistence. Note that one of the trajectories ended up with fully crystallized (albeit highly defected) system, but the probability of this will get smaller with increasing system size.

Each of the simulation trajectories (100 in total) used in this manuscript ran for significantly less than 24 hours. No fine-tuning of the initial packing fraction was required after the initial guess based on known nucleation rates. Hence, the lack of reported observations of this phenomenon is most likely not induced by computational cost. Instead, it is likely the result of the fact that most simulation studies fall in one of three camps: (i) ones that aim to study the properties of the bulk phases (and hence avoid the regime where nucleation occurs), (ii) ones that aim to study the nucleation process (and hence ignore what happens once the nucleus spans the simulation box), and (iii) ones which aim to study interfaces (and hence prefer to explicitly set up the coexistence to control the orientation of the crystal). As these simulations show, if the aim is to observe a simulation which spontaneously reaches a coexisting state from a disordered fluid, this is readily achievable for monodisperse hard spheres.

\section{Data availability and reproducibility statement}

The data underlying the plots and snapshots in this comment are shared as Supplemental Material. The simulation code was based on the one published at \href{https://github.com/FSmallenburg/EDMD/}{https://github.com/FSmallenburg/EDMD/}, which was only slightly altered to calculate the degree of crystallinity during the run.

\bibliography{refs}

@article{smallenburg2022efficient,
  title={Efficient event-driven simulations of hard spheres},
  author={Smallenburg, Frank},
  journal={Eur. Phys. J. E},
  volume={45},
  number={3},
  pages={22},
  year={2022},
  publisher={Springer}
}

@article{wang2026elusive,
  title={The elusive fluid-and-crystal coexistence state in simulations of monodisperse, hard-sphere colloids},
  author={Wang, J Galen and Dhumal, Umesh and Zakhari, Monica EA and Zia, Roseanna N},
  journal={AIChE Journal},
  volume={72},
  pages={e70275},
  year={2026},
  publisher={Wiley Online Library}
}

@article{wohler2022hard,
  title={Hard sphere crystal nucleation rates: Reconciliation of simulation and experiment},
  author={W{\"o}hler, Wilkin and Schilling, Tanja},
  journal={Phys. Rev. Lett.},
  volume={128},
  number={23},
  pages={238001},
  year={2022},
  publisher={APS}
}

@article{de2025local,
  title={Local composition fluctuations act as precursors for crystal nucleation in polydisperse hard spheres},
  author={de Jager, Marjolein and Castagn{\`e}de, Antoine and Smallenburg, Frank and Filion, Laura},
  journal={arXiv preprint arXiv:2512.20300},
  year={2025}
}

@article{smallenburg2024simple,
  title={A simple and accurate method to determine fluid--crystal phase boundaries from direct coexistence simulations},
  author={Smallenburg, Frank and Del Monte, Giovanni and de Jager, Marjolein and Filion, Laura},
  journal={J. Chem. Phys.},
  volume={160},
  pages={224109},
  number={22},
  year={2024},
  publisher={AIP Publishing}
}

@article{macdowell2006nucleation,
  title={Nucleation and cavitation of spherical, cylindrical, and slablike droplets and bubbles in small systems},
  author={MacDowell, Luis G and Shen, Vincent K and Errington, Jeffrey R},
  journal={J. Chem. Phys.},
  volume={125},
  pages={034705},
  number={3},
  year={2006},
  publisher={AIP Publishing}
}

@article{ten1995numerical,
  title={Numerical evidence for bcc ordering at the surface of a critical fcc nucleus},
  author={Ten Wolde, Pieter Rein and Ruiz-Montero, Maria J and Frenkel, Daan},
  journal={Phys. Rev. Lett.},
  volume={75},
  number={14},
  pages={2714},
  year={1995},
  publisher={APS}
}

@article{van2012parameter,
  title={A parameter-free, solid-angle based, nearest-neighbor algorithm},
  author={Van Meel, Jacobus A and Filion, Laura and Valeriani, Chantal and Frenkel, Daan},
  journal={J. Chem. Phys.},
  pages={234107},
  volume={136},
  number={23},
  year={2012},
  publisher={AIP Publishing}
}

@article{lechner2008accurate,
  title={Accurate determination of crystal structures based on averaged local bond order parameters},
  author={Lechner, Wolfgang and Dellago, Christoph},
  journal={J. Chem. Phys.},
  volume={129},
  number={11},
  pages={114707},
  year={2008},
  publisher={AIP Publishing}
}

\end{document}